\documentclass[12pt]{article} 
\usepackage[dvips]{graphics}
\title{Quantum Fields and Translations}  
\author{{\it Richard Shurtleff~}\thanks{affiliation and mailing 
address: Department of Applied Mathematics and Sciences, 
Wentworth Institute of Technology, 550 Huntington Avenue, 
Boston, MA, USA, ZIP 02115, telephone number: (617) 989-4338, e-mail address: shurtleffr@wit.edu}} 
\begin{document} 

\maketitle

\begin{abstract} 

 Modifications of a free quantum field calculation using translation-related concepts and general translation representations yield quantum fields for massive particles that as a consequence follow the classical trajectories of electrodynamics and geometrodynamics. The work allows an explanation for the unexpectedly high energy of cosmic rays. The explanation can be tested at the Large Hadron Collider once high energy proton beams are operational. 



\end{abstract}

\section{Introduction}

Combinations of rotations, boosts and translations are symmetry transformations that preserve the scalar products of Minkowski coordinate differences in flat spacetime.  Quantum fields and particle states are among the objects that are required to transform by linear representations (reps) of these symmetry transformations.  

Consider one way that symmetries can be useful. When linear combinations of objects are required to transform differently than the objects themselves, the coefficients of such linear combinations are constrained. The coefficients are not always constrained to vanish. Clebsch-Gordon coefficients can be determined by such constraints.\cite{CGcoeff} Another example is a derivation of free quantum fields that by using such constraints succeeds in grounding free quantum fields on the principles of special relativity.\cite{W} In this article we modify the latter example.

The quantum field is required to transform with a nonunitary rep, while being a linear combination of annihilation and creation operators. The operators transform with a unitary rep. The contrasting reps, nonunitary versus unitary, determine the coefficients of the linear combination and, therefore, determine the quantum field. 

\vspace{0.2cm}

\noindent The modifications to the free field calculation(FFC):

\noindent 1. In the FFC, the quantum field is invariant under translation, which is the trivial rep of translations. In this article, the quantum field transforms by a nonunitary rep of the Poincar\'{e} group that may include nontrivial representations of translations.

\noindent 2. In the FFC, the momentum is invariant under translation; again the trivial translation rep.  In this article, the most general Poincar\'{e} rep for four-vectors is used, leading to a notion of `parallel translation' for four-vectors defined in Sec. \ref{Phi}.

\noindent 3. The following possibility is not considered in the FFC. After a translation, there is no evidence among the coordinate differences just what displacement is applied. It is inferred that no measurable consequences should follow if one displacement is applied to the quantum field and a second, independent displacement is applied to the annihilation and creation operators and the particle state. 

\noindent 4.  The following possibility is not considered in the FFC. While not related to translations and somewhat technical, note that the quantum field is a linear combination of operators and not a sum over the particle states that those operators add or remove. Furthermore an operator adds or removes the same state no matter where in spacetime the quantum field is being constructed. Location is not transferred in the construction. We infer that the reference frame for the particle states could differ from the reference frame for the quantum field. We allow the two frames to differ, thereby introducing an arbitrary Lorentz transformation. 

\vspace{0.2cm}

We find that the modifications 2 and 3 are just what is needed to obtain the particle motion found in electrodynamics and geometrodynamics. Modification 2 is closely related to gauge theory since it involves parallel translation. Modification 3 is new, allowing flexibility in the choice of a displacement for the operators and particle state and that choice is constrained by the contrasting transformation requirements of the quantum fields and operators. Introducing an explicit interaction is avoided, just as in the FFC but now with forces. 

While the motion is the standard electrodynamics and geometrodynamics, there are new effects due to the second displacement allowed for the operators and particle states. The effects may be evident for sufficiently energetic particles even in weak gravitational fields. Looking for unusual phenomena in this limit, one finds cosmic rays. Intriguingly, cosmic rays are observed to have much more energy than they are expected to have if they acquire their energy in this galaxy.\cite{puzzle} The new effects just mentioned can explain the unexpectedly high energies as a local phenomenon related to the local gravitational field. See Fig. 1. The effect is unrelated to the negligible acceleration due to traditional gravitational effects. 

Coincidentally, the energy region where the new effect would become observable happens to be the region that the Large Hadron Collider is preparing to explore.\cite{LHC} If the effect is real, then it should be apparent at the LHC. The beam would travel and be deflected by the traditional electrodynamics, but upon collision, the proton should deliver much more energy than the trajectory energy. A proton in a 7 TeV trajectory would contribute 23.5 TeV to a collision.

Since the calculation is based on transformation reps and does not introduce an explicit interaction, the 16.5 TeV difference between the trajectory and effective state energies seems to appear from nowhere. It may be that there is an associated interaction, perhaps updating a version of current gravitation theory, that can identify an energy source for the energy difference. Should the effect be observed at the LHC, there would then be some experimental results to assist in the search for the presumed energy source. 

The basis for the effect is that there are two phases in the theory, each the scalar product of a displacement and a momentum, and each with its own momentum. See modification 3 listed above. The trajectory momentum is the same as in traditional electrodynamics and geometrodynamics. But there is a separate, effective particle state momentum that differs from the trajectory momentum at high energies in a weak gravitational field. This effect may be present but may not be noticed at lower energies in the weak field of the Earth.  

Sec. \ref{Phi} gives the general nonunitary translations available to four-vectors and parallel translation of four-vectors is defined. In Sec. \ref{states} the unitary transformations of particle states are discussed. In Sec. \ref{Qfields} a quantum field is constructed as a sum over the creation and annihilation operators of the particle states. The calculation builds on the construction of free fields by Weinberg.\cite{W} Also in Sec. \ref{Qfields} the fact that coordinate differences are invariant under translations leads to two phases, one for the coefficients and one for the particle states and operators. In Sec. \ref{ClassLimit} one of the phases is shown to imply classical trajectories in electromagnetic and gravitational fields. In Sec. \ref{s3} the other phase is adjusted to explain the observation of high energy cosmic rays. The same effect that explains cosmic rays may be observable at the Large Hadron Collider which is undergoing preparations to begin runs as this is written. The predicted effect at the LHC is discussed in Sec. \ref{LHC}.

\pagebreak
\section{Parallel Translation of a Four-Vector } \label{Phi}

Among the objects invariant under translations is the difference of the Minkowski coordinates for two events. The Poincar\'{e} group is the set of transformations leaving invariant the scalar product of coordinate differences. Among these transformations, translations are unique in preserving the differences themselves. Any translation leaves all coordinate differences unchanged because the displacement cancels out,
\begin{equation} \label{deltaX}
     (x_{2} + b) - (x_{1} + b) = x_{2} - x_{1} \, .
\end{equation}
The `Lorentz group' herein indicates the set of successive rotations and boosts (no translations) that, except for the identity, change coordinate differences but preserve their scalar products.

Coordinate differences are not the only translation invariant quantities; also invariant are objects that transform by irreducible representations of the Lorentz group, i.e. those objects with spin $(A,B).$  Minkowski coordinate differences are four-vectors, transforming with spin $(1/2,1/2),$ an irreducible Lorentz rep.

Reducible Lorentz reps that transform with spin $(A,B) \oplus (A\pm 1/2, B \pm 1/2)$ are said to be {\it{linked}}. Linked objects and only linked objects can change with translations.\cite{vectormomentumMATS} Thus, by linking four-vectors, spin $(1/2,1/2),$ with objects of spin $(0,0)$ or $(0,1)$ or $(1,0)$ or $(1,1)$ one can have four-vectors that change upon translation. In texts one often sees a five component object that links a four-vector with a scalar, spin $(0,0),$ to illustrate the effects of translation.\cite{Tung5D}

The spins $(0,0)$ and $(0,1)$ and $(1,0)$ and $(1,1)$ form the spin composition of second rank tensors. Thus we combine the four components of a four-vector $v$ with the sixteen components of a second rank tensor $T$ to create a multicomponent object $\Phi,$ 
\begin{equation} \label{pT1}
 \Phi_{m}  = \pmatrix{v^{\alpha} \cr T^{\gamma \delta}_{(\Phi)} } \, ,
\end{equation}
where the index $m$ runs from 1 to 20 in a fixed sequence over the indices $\alpha,\gamma,\delta$ that each run from 1 to 4, indicating Minkowski coordinates with $x^{4}$ = $x^{t}$ as the time component. As part of the multicomponent object $\Phi,$ the four-vector $v$ has the most general Poincar\'{e} transformations.

Representations for transforming four-vectors and tensors are well known. We select the angular momentum and boost generators to be 
\begin{equation} \label{J11} 
 (J^{\rho \sigma})^{\mu}_{\nu}  = i \pmatrix{ \eta^{\sigma \mu} \delta^{\rho}_{\nu} - \eta^{\rho \mu} \delta^{\sigma}_{\nu}  && 0 \cr 0 && - \eta^{\rho \gamma} \delta^{\sigma}_{\epsilon} \delta^{\delta}_{\xi}+ \eta^{\sigma \gamma} \delta^{\rho}_{\epsilon} \delta^{\delta}_{\xi} - \eta^{\rho \delta} \delta^{\sigma}_{\xi} \delta^{\gamma}_{\epsilon}+ \eta^{\sigma \delta} \delta^{\rho}_{\xi} \delta^{\gamma}_{\epsilon}  }\, ,
\end{equation} 
where $\delta$ is one for equal indices and zero otherwise and $\eta$ is the spacetime metric, \linebreak $\eta$ =  diag$\{+1,+1,+1,-1\}.$ The matrices in (\ref{J11}) can be found in the literature.\cite{WeinbergV}

Less well-known are the four $20 \times 20$ momentum matrices, $P_{(\Phi)}^{\mu}.$ We use 
$$ 
 (P_{(\Phi)}^{\sigma})^{\alpha}_{\gamma \delta}  = i \pmatrix{ 0 && \pi_{1} \delta^{\sigma}_{\gamma} \delta^{\alpha}_{\delta}  + 
 \pi_{2}\delta^{\sigma}_{\delta} \delta^{\alpha}_{\gamma}  +  \pi_{3} \eta^{\sigma \alpha} \eta_{\gamma \delta}
  + \pi_{4}  \eta^{\sigma \rho} \eta^{\alpha \kappa} \epsilon_{\rho \kappa \gamma \delta} \cr 0 && 0  }  \, ,
$$ 
where $\epsilon$ is the antisymmetric symbol and the constants $\pi_{i}$ have dimensions of an inverse distance.\cite{RS} There are two types of momentum matrices, those shown above with only the 12-block nonzero and those with only the 21-block nonzero. The expression for the 21-momentum matrices is similar to the above expression. We need the 12-rep here since the 12-rep changes the four-vector $v$ and leaves the tensor $T$ unchanged.

Any Poincar\'{e} transformation can be written as a homogeneous Lorentz transformation $\Lambda$ followed by a translation through a displacement $\delta x.$ For the transformation of $\Phi$ we have, 
\begin{equation} \label{transf2}
 \Phi_{l}^{\prime}  = \sum_{\bar{l}} D^{(\Phi)}_{l \bar{l}}(\Lambda,\delta x)  \Phi_{\bar{l}} =  \sum_{\bar{l}} \left[ \exp{(- i \delta x_{\sigma} P_{(\Phi)}^{\sigma})}\exp{( i \omega_{\mu \nu} J_{(\Phi)}^{\mu \nu}/2)} \right]_{l \bar{l}}  \Phi_{\bar{l}} \, ,
\end{equation}
where $D^{(\Phi)}(\Lambda,\delta x)$ is the matrix representing the Poincar\'{e} transformation $(\Lambda,\delta x)$ and the parameters $\omega_{\mu \nu}$ produce a representation of $\Lambda.$ 

To find the effect of translation on the four-vector $v$ in $\Phi,$ Eq. (\ref{pT1}), let the Lorentz transformation $\Lambda$ be the identity by making the parameters $\omega_{\mu \nu}$ vanish, $\omega_{\mu \nu}$ = 0. Then a translation along the displacement $\delta x$ changes the four-vector $v$ to $v^{\prime}$ by the rule
\begin{equation} \label{Dp2}
{v^{\prime}}^{\alpha}  = v^{\alpha}+ \eta_{\sigma \mu} {T}^{\alpha \sigma }_{(v)}  \delta x^{\mu}  \, ,
\end{equation}
where
$${T}^{\alpha \sigma }_{(v)} \equiv -i  (P^{\sigma}_{12})^{\alpha}_{\beta \gamma} T^{\beta \gamma}_{(\Phi)}
  \, . $$
For a displacement $\delta x$ that is sufficiently small, the translated four-vector $v^{\prime}$ is uniquely determined. For a large displacement one may construct many sequences of sufficiently small displacements. In general, the translated four-vector $v^{\prime}$ depends on the particular sequence selected. This is something like path dependence, but since each sequence of displacements translates all of spacetime, there are infinitely many similar paths for each sequence.

{\it{Parallel Translation of a Four-Vector.  The post-translation four-vector $v^{\prime}$ is {\it{equivalent to}} the pre-translation four-vector $v.$}} 

For each choice of $T$ there is a different parallel translation. This much like parallel transport in affine spaces where the change in $v$ depends on a connection. Here the transport is inhomogeneous in $v,$ so the same term is added to any $v$ even the null four-vector.

The parallel transport of the momentum four-vector has the effect of an implicit hamiltonian since the momentum and energy change upon translation. Since $T$ is an arbitrary second rank tensor, there is much flexibility in the parallel transport rule (\ref{Dp2}) and in the resultant effects. However the process of constructing a quantum field in Sec. \ref{Qfields} and \ref{ClassLimit} constrains the tensor $T$ and so constrains the implicit interaction.

\section{Unitary Transformations of Particle States} \label{states}

Unitary representations (reps) of the Poincar\'{e} group must be infinite dimensional. So, to have finite dimensional square matrix reps, a unitary rep has these matrices dependent on some continuous parameter, so the rep is infinite dimensional. For particle states it is convenient to have that quantity be the momentum; the unitary reps below have matrices that are momentum dependent. Unitary translation reps are closely related to the ubiquitous plane wave.

Let $\Psi_{\sigma}(y)$ indicate the state of a particle with spin $j$ at the event $y$ in the Minkowki coordinates of a given reference frame of spacetime. By basic quantum principles, the reps of the Poincar\'{e} transformations for $\Psi$ are unitary in order to preserve the inner products between states. By the Poincar\'{e} algebra, momentum operators commute so one can expand a state $\Psi_{\sigma}(y)$ over the set of momentum eigenstates $\Psi_{p,\sigma}(y).$

Let $\Psi_{p,\sigma}$ be a single particle eigenstate with eigenmomentum $p$,
\begin{equation} \label{eigen}
     P^{\mu}_{(\Psi)} \Psi_{p,\sigma} = p^{\mu} \Psi_{p,\sigma} \, ,
\end{equation}
where $\sigma$ is the $z$-component of spin. We choose the unitary rep of a given Poincar\'{e} transformation $(\Lambda,\epsilon)$ so that the state changes by the rule \cite{Wp68VolI}
\begin{equation} \label{Ptrans1}
     U(\Lambda, \epsilon) \Psi_{p,\sigma} =  e^{-i \Lambda p \cdot \epsilon} \sqrt{\frac{(\Lambda p)^t}{p^t}}  \sum_{\bar{\sigma}} D^{(j)}_{\sigma \bar{\sigma}}(W^{-1}) \Psi_{p,\bar{\sigma}} \, ,
\end{equation}
where $D^{(j)}$ is the rotation matrix for the spin $j$ of the particle and the rotation $W$ is the Wigner rotation given by
\begin{equation} \label{WLp}
W(\Lambda, p) = L^{-1}(\Lambda p)\Lambda L(p) \, ,
\end{equation}
with $L(p) $ a standard transformation taking $k^{\mu}$ = $\{0,0,0,m\}$ to $p.$ We choose $L$ to be
$$L^{i}_{k}(p) = \delta^{i}_{k} + (1+\gamma)^{-1} m^{-2} p^{i}p^{k}\, ,$$
 \begin{equation} \label{L} L^{i}_{4} = L^{4}_{i} = m^{-1}p^{i} \quad {\mathrm{and}} \quad L^{4}_{4} = \gamma = m^{-1} p^{4}\, ,
\end{equation}
where $i,k \in$ $\{1,2,3\}$ = $\{x,y,z\}$ and $m$ is the mass of the particle. Since the mass is the magnitude of the momentum, one can find the energy $p^{t} \geq m$ knowing the three-momentum $\overrightarrow{p}$ = $\{p^{1},p^{2},p^{3}\}.$ 

Note that for a given Poincar\'{e} transformation $(\Lambda, \epsilon),$ the rotation matrix $D^{(j)}$ is different for different momenta and hence infinite dimensional even though the rotation matrices are finite $(2j+1)-$square matrices. 

The particle eigenstates may be added to or removed from a multiparticle state. One can show that the creation and annihilation operators, $a^{\dagger}_{\sigma}(\overrightarrow{p})$ and $a_{\sigma}(\overrightarrow{p})$ respectively, transform with much the same representations as the particle state  $\Psi$ and its adjoint. Thus the transformations of $a^{\dagger}_{\sigma}(\overrightarrow{p})$ and $a_{\sigma}(\overrightarrow{p})$ can be written as unitary $(2j+1)$-dimensional matrices that depend on the momentum $p.$

\section{Quantum Fields; Event-By-Event } \label{Qfields} 

In this section, we work at a single given event. There are Poincar\'{e} transformations, but these only provide the given event with new coordinate labels. The parallel translation of momentum or other four-vectors by (\ref{Dp2}) is reserved for active translations that map events onto other events. The momenta in this section change with rotations and boosts to maintain scalar products with other four-vectors, but the momenta do not change with translations.  

At each event $x$ in spacetime a value for a quantum field may be constructed as a sum of annihilation and creation operators with invariant coefficients.\cite{W} It should be clear that working at a given event $x$ in no way confines the single particle states $\Psi_{\sigma}$ added or removed by the operators. The field is a sum over operators, not states, and the operators add or remove the same states at whatever event the field is constructed. The operators are the same at all events in spacetime. 

While both particle states and quantum fields are defined over all of spacetime, it may be that they are defined in reference frames that are related by some transformation $\lambda.$ Since the field is a sum over the operators and not a sum over the states, it may be that the field is defined in one reference frame and creates or removes states in a second reference frame. Since any two reference frames are related by some Lorentz transformation,  there is a Lorentz transformation $\lambda$ transforming the frame of the states to the frame for the field,
\begin{equation} \label{yTOx}
 x  = \lambda y \, .
\end{equation}
This seemingly innocuous property is needed to explain cosmic rays in Sec. \ref{s3}. 

Let the given event have coordinates $x$ in an initial reference frame. The quantum field $\psi_{l}(x)$ can be separated into linear combinations of annihilation and creation operators, $\psi_{l}(x)$ = $\kappa \psi^{+}_{l}(x)$ + $\lambda \psi^{-}_{l}(x).$ One has an annihilation field $\psi^{+}$ and a creation field $\psi^{-}$ given by 
$$\psi^{+}_{l}(x) = \sum_{\sigma} \int d^3 p \enspace u_{l\sigma}(x,{\overrightarrow{p}}) a_{\sigma}({\overrightarrow{p}})  \, ,$$
\begin{equation} \label{psi+}
\psi^{-}_{l}(x) = \sum_{\sigma} \int d^3 p \enspace v_{l\sigma}(x,{\overrightarrow{p}}) a^{\dagger}_{\sigma}({\overrightarrow{p}})  \, .
\end{equation}
Since the fields are linear combinations of operators, quantum fields are also operators. We reserve the term `operator' for the annihilation and creation operators. 

The coefficients $u$ and $v$ can be determined from the ways the fields and operators transform to preserve spacetime symmetries: ({\it{i}}) the operators $a$ and $a^{\dagger}$ transform under Poincar\'{e} transformations with a unitary representation (rep), ({\it{ii}}) the coefficients are required to be invariant and ({\it{iii}}) the quantum field transforms by a nonunitary rep. Much as Clebsch-Gordon coefficients connect quantities that transform differently under rotations, the coefficients $u$ and $v$ connect ({\it{i}}) the operators $a$ and $a^{\dagger}$ and ({\it{iii}}) the fields $\psi^{\pm}_{l}$. As with Clebsch-Gordon coefficients, the different transformation regimens constrain the coefficients so much that the coefficients are essentially determined. 

Before turning to the calculation, a translation-related property needs to be introduced. The symmetries of spacetime are built on the behavior of coordinate differences. Under rotations and boosts, the coordinate differences change, but in ways that preserve spacetime scalar products. Translations are different. By (\ref{deltaX}), all translations preserve the coordinate differences themselves because any displacement cancels upon subtracting the coordinates. By comparing the coordinate differences after two translations have been applied it is impossible to determine if the displacements were the same or different. 

It is assumed in other such calculations that the displacement applied to operators is identical to the displacement applied to the coordinates of the given spacetime event $x.$ However, all coordinate differences are unchanged by any translation. Here we apply a possibly different displacement $\epsilon$ to the operators than the displacement $\delta x$ that is applied to the fields. 

{\it{Field and Operator Displacements. Suppose the Minkowski coordinates $x$ of the given event transform with a combination $\Lambda$ of rotations and boosts followed by a translation through a displacement $b,$ in symbols: $x \rightarrow$ $\Lambda x + b.$ Then the annihilation and creation operators must transform with the same Lorentz transformation $\Lambda$ but this is followed by a displacement $\epsilon$ that is a suitably differentiable, coordinate-dependent function $\epsilon (\Lambda,x,b)$ of the Poincar\'{e} transformation $(\Lambda,b).$ }}

The annihilation and creation operators transform by ({\it{i}}) a unitary representation of the Poincar\'{e} transformation $(\Lambda,\epsilon)$ in much the same way as a particle state, (\ref{Ptrans1}). We have
$$U(\Lambda,b) \psi^{+}_{l}(x) {U}^{-1}(\Lambda,b) = \sum_{\sigma} \int d^3 p \enspace u_{l\sigma}(x,{\overrightarrow{p}})e^{i \Lambda p \cdot \epsilon(\Lambda,x,b)} \sqrt{\frac{(\Lambda p)^t}{p^t}}  \sum_{\bar{\sigma}} D^{(j)}_{\sigma \bar{\sigma}}(W^{-1})  a_{\bar{\sigma}}({\overrightarrow{\Lambda p}}) \, ,$$
\begin{equation} \label{Da+}
U(\Lambda,b) \psi^{-}_{l}(x) {U}^{-1}(\Lambda,b) = \sum_{\sigma} \int d^3 p \enspace v_{l\sigma}(x,{\overrightarrow{p}})e^{-i \Lambda p \cdot \epsilon(\Lambda,x,b)} \sqrt{\frac{(\Lambda p)^t}{p^t}}  \sum_{\bar{\sigma}} D^{(j)\ast}_{\sigma \bar{\sigma}}(W^{-1})  a^{\dagger}_{\bar{\sigma}}({\overrightarrow{\Lambda p}}) \, .
\end{equation}
Note that the coefficients $u$ and $v$ remain ({\it{ii}}) invariant. The labels ({\it{i}}) and ({\it{ii}}) refer to the transformation regimens listed just after Eq. (\ref{psi+}).

The unitary transformation $U(\Lambda,b)$ is required to have the effect of ({\it{iii}}) a nonunitary transformation on the fields. One requires that 
\begin{equation} \label{Dpsi}
U(\Lambda,b) \psi^{\pm}_{l}(x) {U}^{-1}(\Lambda,b) = \sum_{\bar{l}} D^{-1}_{l \bar{l}}(\Lambda,b)  \psi^{\pm}_{\bar{l}}(\Lambda x + b) \, ,
\end{equation}
where $\Lambda x + b$ are the transformed coordinates of the given event and $D(\Lambda,b)$ is the nonunitary matrix representing the Poincar\'{e} transformation $(\Lambda,b).$ The rep $D(\Lambda,b)$ corresponds to the spin composition of the field $\psi$ which we denote by  $(A,B)\oplus(C,D)\oplus \ldots \,.$ We allow a reducible Lorentz spin composition so that the spins may be linked and nontrivial reps of translation can occur. 

 By (\ref{psi+}), (\ref{Da+}) and (\ref{Dpsi}) one finds that the $u$s and $v$s at $x$ and $\Lambda x +b$ depend on the $u$s and $v$s at the origin as follows,
 $$   e^{+i \Lambda p \cdot [\epsilon(\Lambda,x,b)-\Lambda \epsilon(1,x,-x) +  \epsilon(1, \Lambda x + b,-\Lambda x - b)] }\sum_{\bar{l}} D_{l \bar{l}}(\Lambda,0) u_{\bar{l}\sigma}(0,{\overrightarrow{p}}) = \hspace{5cm}$$ $$\hspace{5cm} \sqrt{\frac{ (\Lambda p)^t}{p^t}} \sum_{\bar{\sigma}} u_{l\bar{\sigma}}(0,{\overrightarrow{\Lambda p}}) D^{(j)}_{\bar{\sigma} \sigma}(W(\Lambda,p))    \, $$
$$e^{-i \Lambda p \cdot [\epsilon(\Lambda,x,b)-\Lambda \epsilon(1,x,-x) +  \epsilon(1,\Lambda x + b,-\Lambda x - b)]}\sum_{\bar{l}} D_{l \bar{l}}(\Lambda,0) v_{\bar{l}\sigma}(0,{\overrightarrow{p}}) = \hspace{5cm}$$ 
\begin{equation} \label{Du4}
     \hspace{5cm} \sqrt{\frac{ (\Lambda p)^t}{p^t}} \sum_{\bar{\sigma}} v_{l\bar{\sigma}}(0,{\overrightarrow{\Lambda p}}) {D^{(j)}}^{\ast}_{\bar{\sigma} \sigma}(W(\Lambda,p))   
\end{equation}
Note that the right-hand-sides of Eq. (\ref{Du4}) do not depend on the coordinates $x$ of the event in the initial reference frame. Nor do they depend on the displacement $b,$ so the left-hand-sides cannot depend on $x$ or $b$ either. 

Since $x$ and $b$ occur only on the left and there only in the operator displacement function $\epsilon,$ one seeks a suitable function $\epsilon(\Lambda,x,b)$ such that the expression in brackets in the phase doesn't depend on $x$ or $b.$ One can show that $\epsilon(\Lambda,x,b)$ must be in the following form,
\begin{equation} \label{epsilon1}
 \epsilon^{\mu}(\Lambda,x,b) =  \epsilon^{\mu}(\Lambda) - \Lambda^{\mu}_{\sigma} [M(x)]^{\sigma}_{\nu} x^{\nu} + [M(\Lambda x+b)]^{\mu}_{\nu} (\Lambda x+b)^{\nu} \, ,
\end{equation}
where $M(x)$ is an arbitrary second rank tensor field defined over the collection of allowed coordinates, i.e.  $\Lambda x+b$ for any $\Lambda$ and $b,$ for the given event. 

For simplicity, we drop the $x-$ and $b-$independent displacement, $\epsilon^{\mu}(\Lambda)$ = 0. One recovers $\epsilon$ = $b$ when the field $M$ is the identity, $M^{\mu}_{\nu}$ = $\delta^{\mu}_{\nu}.$

By (\ref{psi+}) to (\ref{epsilon1}), one finds expressions for the coefficients $u$ and $v$ in terms of the coefficients at the origin with the momentum of a particle at rest,
$$   u_{l\sigma}(x ,{\overrightarrow{p}}) = \sqrt{\frac{ m}{p^{\, t}}} \, e^{i p \cdot Mx} \sum_{\bar{l}} D_{l \bar{l}}(L,x) u_{\bar{l}\sigma}(0,{\overrightarrow{0}})     \, ,$$
\begin{equation} \label{Du8}
    v_{l\sigma}(x ,{\overrightarrow{p}}) =  (-1)^{j+\sigma} \sqrt{\frac{ m}{p^{\, t}}} \, e^{-i p \cdot  Mx} \sum_{\bar{l}} D_{l \bar{l}}(L,x) v_{\bar{l}\sigma}(0,{\overrightarrow{0}})     \, .
\end{equation}
The expressions (\ref{Du8}) along with Eq. (\ref{psi+}) for $\psi^{+}$ and $\psi^{-}$ determine the quantum field $\psi_{l}(x)$ = $\kappa \psi^{+}_{l}(x)$ + $\lambda \psi^{-}_{l}(x).$ 

In the special case with $M$ equal to the identity, the quantum fields found here are the free fields constructed by Weinberg.\cite{W,S1} 

In this section the spacetime symmetry transformations of the various quantities occurs at a single given event and the value of a quantum field at the given event is found. The next section shows how the active translations considered in Sec. \ref{Phi} produce quantum fields that respond to forces.


\section{Dynamics and Classical Limit } \label{ClassLimit}

In this section, the classical time-like path is found using the phase  $p \cdot  Mx$  determined in Sec.~\ref{Qfields}. Also we parallel translate the momentum four-vector as in Sec. \ref{Phi}. Since the momentum is changed upon translation by a term dependent on $T,$ the tensor $T$ represents implicit forces.  One finds that the presence of the tensor $M$ in the phase $p \cdot  Mx$ and the requirement of constant mass leads to an identification of the implicit forces with those of electromagnetism and gravity. 

When the wavelength scale is much smaller than the scale of the motion, it is sometimes sufficiently accurate to consider the quantum field nonzero only in a sequence of four-dimensional regions of spacetime each combining a tubular region of space with a finite interval of time. Heuristically, a particle may be confined by allowing it a suitably wide range of momenta. With the particle confined, the translations of all spacetime can be localized to translations of the region of spacetime containing the particle. So a classical path  for our purposes indicates a sequence of translations applied to all of the relevant portion of spacetime inside a `tubular' trajectory.

By (\ref{Du8}), the quantum fields here are sums of `plane waves' in the form $\exp{(\pm i p \cdot M x)},$ 
where $M,$ which is introduced in (\ref{epsilon1}), is a second rank tensor field. We are thereby lead to assume that the amplitude is proportional to $\exp{( \pm i p \cdot M \delta  x)}$ for a particle in an eigenstate with momentum $p$ to translate through a displacement $\delta x.$  To be well-defined, $\delta x$ must be so small that neither $p$ nor $M$ change significantly along $\delta x.$ So-called polarization effects that result from the mixing of quantum field components are ignored; we focus on the phase.

Let $\exp{(\pm i \Theta)}$ be proportional to the amplitude for the particle in a particular momentum eigenstate to follow a given sequence of sufficiently short translations that bring an event 1 to an event 2 along a trajectory $C.$ Then $\Theta$ is the change of phase along $C$ from event 1 to event 2,  
\begin{equation} \label{I1}
     \Theta =  \int_{1}^{2} p \cdot M dx = \int_{1}^{2}\eta_{\alpha \beta} p^{\alpha} M^{\beta}_{\sigma} dx^{\sigma}   \, ,
\end{equation}
where $\eta$ is the (flat) spacetime metric, $\eta$ = diag$\{1,1,1,-1\}.$  
As discussed above, it is the sequence of translations $\delta x_{a}, \delta x_{b}, \ldots$ that is important, events 1 and 2 must be somewhere in the relevant region of spacetime containing the particle. It is assumed that $M$ changes slowly over that region so that the same $\Theta$ occurs with any of the allowed choices for events 1 and 2.

Now we find timelike trajectories with extreme phase. Consider the phase shift $\delta \Theta$ = $p \cdot M \delta x$ over a displacement $\delta x$ short enough that $M$ varies negligibly along $\delta x$ in the region where the particle is confined.  One finds that the extreme phase shift, $\delta \Theta$ = $-m \delta \tau,$ occurs for timelike $M\delta x$ when $M\delta x$ is proportional to $p,$
\begin{equation} \label{pMdx2}  M^{\alpha}_{\mu} {\delta X}^{\mu} = m^{-1} p^{\alpha} \delta \tau   \, .
 \end{equation}
Let upper case letters, as in $\delta X,$ stand for displacements $\delta x$ that make $\delta \Theta$ extreme. Assuming, for simplicity, that $M$ has an inverse, then the collection of such trajectories for the various momenta $p$ determines a coordinate system.

 By indicating the derivative with respect to $\tau$ with a dot,   i.e. $\dot{X} \equiv$  $dX/d \tau,$ one can rewrite (\ref{pMdx2}) in a traditional notation,
\begin{equation} \label{Ydot1}   p^{\alpha} = m M^{\alpha}_{\mu} \dot{ X}^{\mu}\, .
 \end{equation}
Having discussed the classical trajectories, we turn now to the requirement that the mass be constant.

With the Wigner class of momenta for a massive particle, the (flat) spacetime magnitude of the momentum is the particle mass, 
\begin{equation} \label{m2}  \eta_{\alpha \beta} p^{\alpha}p^{\beta} = -m^{2} \, ,
\end{equation}
where $m$ is the particle mass, a constant. It is also required that $p^{t}\geq m >$ 0.  
It follows from (\ref{Ydot1}) and (\ref{m2}) that the `curved spacetime magnitude'  of $\dot{X}$ is constant,
\begin{equation} \label{g2}  g_{\mu \nu} \dot{ X}^{\mu}\dot{ X}^{\nu} = -1    \, ,
\end{equation}
where the `curved spacetime metric' $g_{\mu \nu}$ is defined by
\begin{equation} \label{g3}  g_{\mu \nu} \equiv  \eta_{\alpha \beta} M^{\alpha}_{\mu}M^{\beta}_{\nu} \, .
\end{equation}
Introduced in Eq. (\ref{epsilon1}), $M$ is an arbitrary field, so $g_{\mu \nu}$ is an arbitrary field. Assuming, as previously noted, that $M$ has an inverse, then $g_{\mu \nu}$ has an inverse that we write with raised indices, $g^{\rho \sigma}.$ Since the flat spacetime metric $\eta_{\alpha \beta}$ is symmetric, both $g_{\mu \nu}$ and $g^{\rho \sigma}$ are symmetric.

Having derived (\ref{g2}) and since we are borrowing terms from general relativity, we are justified in calling the quantity $d\tau$ = $\sqrt{-g_{\mu \nu} { dX}^{\mu}{d X}^{\nu}} $ {\it{the `proper time' associated with the curved spacetime metric $g_{\mu\nu}$ along a trajectory of extreme phase.}}

We now introduce forces by assuming that the momentum is parallel translated, see Eq. (\ref{Dp2}). The parallel translation of momentum induces a parallel translation of momentum eigenstates as follows. The collection of momentum eigenstates is the same before and after a translation along a displacement $\delta x$ because the momentum operator $P^{\mu}_{(\Psi)}$ in (\ref{eigen}) is invariant under translation. Now define the pre-translation eigenstate $\Psi_{p,\sigma}$ in the collection of momentum eigenstates to be {\it{equivalent}} to the post-translation eigenstate $\Psi_{p^{\prime},\sigma}$ in the collection when $p^{\prime}$ is equivalent to $p.$ 

The Dynamical Postulate can be stated as if for free fields: 

\noindent{\it{A particle in a given eigenstate remains in equivalent eigenstates as spacetime is translated.}}

By the Dynamical Postulate, the translation rule (\ref{Dp2}) with momentum $p$ as four-vector $v,$ $v \rightarrow$ $p,$ applies along a trajectory of extreme phase. Now define $T_{(p)}$ by
 $$  T^{\alpha \mu }_{(p)} \equiv g^{\mu \nu}\eta_{\sigma \nu} {T}^{\alpha \sigma }_{(v)}     \, .$$
Then, by (\ref{Dp2}) with $v \rightarrow$ $p$ and the displacement $\delta x$ along the extreme phase displacement, i.e. $\delta x$ = $\delta X,$ one finds
\begin{equation} \label{Dp3a}
\dot{ p}^{\alpha} =  g_{\sigma \mu} T^{\alpha \sigma }_{(p)}  {\dot{X}}^{\mu}  \, .
\end{equation}
This equation relates the change of momentum $p$ with respect to proper time to the change in the coordinates $X$ of particle trajectories with respect to proper time.

Now we combine the relation (\ref{Ydot1}) based on constant mass with the parallel translation-based Eq. (\ref{Dp3a}). One finds an equation for the `four-acceleration' $\ddot{X}$ by defining  the quantity $T^{\mu \sigma }_{(x)}$ as
 $$ T^{\mu \sigma }_{(x)} \equiv m^{-1} (M^{-1})_{\alpha}^{\mu} \, T^{\alpha \sigma }_{(p)} - g^{ \sigma \nu} (M^{-1})_{\alpha}^{\mu}\frac{dM^{\alpha}_{\nu}}{d\tau}    \, .$$
Then, by substituting the expression (\ref{Ydot1}) for $p$ into (\ref{Dp3a}), one finds that
\begin{equation} \label{D2x1}
\ddot{ X}^{\mu} =  g_{\sigma \nu} T^{\mu \sigma }_{(x)}  {\dot{X}}^{\nu}  \, .
\end{equation}

Without loss of generality, one can introduce quantities $F$ and $\Gamma$ so that $T_{(x)}$ takes the form
\begin{equation} \label{Tx2}  T^{\rho \sigma}_{(x)} = \frac{e}{m} F^{\rho \sigma} - g^{\sigma \kappa} \Gamma^{\rho}_{\kappa \lambda} \dot{X}^{\lambda}    \, ,
\end{equation}
with $F$ antisymmetric
$$F^{\sigma \rho } = - F^{\rho \sigma}    \, .$$
By going to a frame in which $\dot{X}$ has only its time component nonzero,  $\dot{X}$ = $\{0,0,0,\dot{X}^{t}\},$ one can see that (\ref{Tx2}) does not constrain $T_{(x)}$ because there are sixteen components of $T^{\mu \nu}_{(x)}$ while there are six free components of the antisymmetric $F^{\sigma \rho }$ and sixteen free components of $\Gamma^{\rho}_{\kappa t}.$

By substituting the expression (\ref{Tx2}) for $T_{(x)}$  in the equation (\ref{D2x1}) for $\ddot{X}$ , one finds that
\begin{equation} \label{D2x2}
\ddot{ X}^{\mu} = \frac{e}{m} g_{\sigma \nu}  F^{\mu \sigma}{\dot{X}}^{\nu} -  \Gamma^{\mu}_{ (\lambda \nu)} \dot{X}^{\lambda} {\dot{X}}^{\nu}  \, .
\end{equation}
This is the equation for the classical trajectory of a particle of mass $m$ and charge $e$ in an electromagnetic field $F^{\rho \sigma}$ and the gravitational field due to a curved spacetime metric $g_{\mu \nu}$ when the symmetric part of $\Gamma,$ i.e. $\Gamma^{\mu}_{ (\lambda \nu)} $ is the Christoffel symbol for the metric $g_{\mu \nu}.$\cite{Somebody}

One can show that $\Gamma$ is the Christoffel symbol for the metric $g_{\mu \nu}$ because the curved spacetime magnitude of $\dot{X}$ is constant along the trajectory of extreme phase by (\ref{g2}). If the derivative is taken with respect to the proper time $\tau,$ i.e. $d(g_{\mu \nu} \dot{ X}^{\mu}\dot{ X}^{\nu})/d\tau$ = 0, and one substitutes (\ref{D2x1}) and $ \dot{g}_{\mu \nu}$ = $\left( \partial{g_{\mu \nu}}/\partial{X^{\lambda}}\right) \dot{X}^{\lambda},$ then one finds    
\begin{equation} \label{DgXX1}  \left( \frac{\partial{g_{\mu \nu}}}{\partial{X^{\lambda}}} - g_{\rho \nu} \Gamma^{\rho}_{(\mu \lambda)}  - g_{\rho \mu} \Gamma^{\rho}_{(\nu \lambda)} \right) \dot{X}^{\lambda} \dot{ X}^{\mu}\dot{ X}^{\nu} = 0    \, ,
\end{equation}
where, by the symmetry evident in the product of the $\dot{X}$s, only the symmetric part of $\Gamma$ contributes. Since the trajectories of extreme phase have sufficiently arbitrary tangents $\dot{X}$, the expression in parentheses in (\ref{DgXX1}) vanishes. 

The vanishing of the expression in parentheses in (\ref{DgXX1}) leads by permuting indices $\lambda, \mu, \nu$ to an expression for the symmetric part of $\Gamma.$ One deduces that
\begin{equation} \label{Gamma1}  \Gamma^{\rho}_{(\mu \nu)} =  \frac{g^{\rho \lambda}}{2} \left( \frac{\partial{g_{\lambda \mu}}}{\partial{X^{\nu}}} +\frac{\partial{g_{\nu \lambda}}}{\partial{X^{\mu}}} - \frac{\partial{g_{\mu \nu}}}{\partial{X^{\lambda}}} \right)
   \, .
\end{equation}
Thus the symmetric part of the quantity $\Gamma$ is indeed the Christoffel connection of the curved metric $g.$\cite{Christoffel}

It should be emphasized that the problem of determining the electromagnetic field $F^{\rho \sigma}$ and the  metric $g_{\mu \nu}$ from the motion of source charges and masses is not considered here. Thus the identification of $F^{\rho \sigma}$ and $g_{\mu \nu}$ with the electromagnetic field and the curved spacetime metric, respectively, is based largely on the resemblance of (\ref{D2x2}) and (\ref{Gamma1}) to equations from general relativity.

One defines the `covariant derivative' of $\dot{X}$ to be
\begin{equation} \label{D2x3}
\frac{D\dot{ X}^{\mu}}{d\tau} \equiv  \ddot{ X}^{\mu} +  \Gamma^{\mu}_{ (\lambda \nu)} \dot{X}^{\lambda} {\dot{X}}^{\nu}  \, .
\end{equation}
Then (\ref{D2x2}) can be written in terms of `covariant' quantities. One has
 \begin{equation} \label{D2x4}
\frac{D\dot{ X}^{\mu}}{d\tau} = \frac{e}{m} g_{\sigma \nu}  F^{\mu \sigma}{\dot{X}}^{\nu}   \, .
\end{equation}
This equation is invariant when the extreme coordinates $X$ are replaced by some other set of extreme coordinates $Z,$ which entails replacing $M$ with some other quantity $N$ in the expression (\ref{epsilon1}) for the displacement $\epsilon$ of the operators. 

Thus general covariance arises from the freedom to choose $M$ in (\ref{epsilon1}) which in turn arises from the arbitrariness assumed in (\ref{Da+}) for the displacement $\epsilon$ when spacetime is displaced by an amount $b.$ The origin of general covariance in this article is the fact that any displacement of (flat) spacetime preserves all coordinate differences.

\section{Cosmic Rays} \label{s3}

In this section we interpret the formula (\ref{epsilon1}) for the displacement $\epsilon$ of the operators. Recall that when the operators are transformed by the Poincar\'{e} transformation $(\Lambda,\epsilon),$ the field is transformed by $(\Lambda,b),$ i.e. with a different translation. Eq. (\ref{epsilon1}) details the dependence of $\epsilon$ on $M,$ $\Lambda,$ $x$ and $b$ needed to construct the field as a sum of operators.

Since the quantum field is a sum over the annihilation and creation operators of particle states and not a sum over the particle states themselves, there is a separation of field and states. In view of this, the particle states $\Psi_{\sigma}(y)$ are allowed to depend on spacetime coordinates $y$ when the field $\psi_{l}(x)$ is defined over the spacetime coordinates $x,$ with $x$ and $y$ possibly in different reference frames. Let the frames be related by the Lorentz transformation $\lambda,$ in symbols: $x$ = $\lambda y,$ as in (\ref{yTOx}). 

While the quantum field is indirectly related to the states, the operators have properties that are directly related to the states they remove or add. When the operators transform by the unitary Poincar\'{e} transformation $(\Lambda,\epsilon),$ the particle states must transform by the same transformation $(\Lambda,\epsilon).$ \cite{Wchapter4Vol1}

Thus, in this section we look at the effects of the formula (\ref{epsilon1}) on the particle states. Since $\epsilon$ depends on $M$ and, by Eq. (\ref{g3}),  $M$ is related to the curved spacetime metric $g_{\mu \nu},$ the effects on the particle states depend on gravity. 

It turns out that the momentum-energy of a particle state may depend on the local gravity. We take advantage of this possibility to obtain an explanation of the ultrahigh energy states of some cosmic rays detected in Earth's gravity. Assumptions are made with the intent of explaining cosmic rays energies and if it happens that the explanation is proven false, then alternate assumptions can be made to recover agreement of the model in this article with experimental results. 

We show that the gravitational dependence of particle state energy can be adjusted so that the ultrahigh energy cosmic rays have much lower energies in interstellar space.

To begin with, let us focus on momentum eigenstates. Since the momentum operator for the transformation of the spacetime coordinates $y$ is $i \partial/ \partial y,$ the eigenstates are plane waves proportional to $\exp{(\pm i p \cdot y)}.$ As before we ignore any polarization effects. Now we try to understand how the plane wave can transform with one displacement $\epsilon_{A}$ for a field at event $A$ and with another displacement $\epsilon_{B}$ for the field at event $B.$

Note that the phase factor at any event, say $A,$ by itself realizes a representation of the Poincar\'{e} group. The successive transformations $(\Lambda_{1},\epsilon^{A}_{1})$ followed by $(\Lambda_{2},\epsilon^{A}_{2})$ applied to the phase factor at the event $A$ yield 
$$  e^{i p_{0} \cdot y^{A}_{0}} \rightarrow e^{i \Lambda_{1}p_{0} \cdot (\Lambda_{1}y^{A}_{0} +\epsilon^{A}_{1})}  \rightarrow e^{i \Lambda_{2}\Lambda_{1}p_{0} \cdot (\Lambda_{2}\Lambda_{1}y^{A}_{0} +\Lambda_{2}\epsilon^{A}_{1}+\epsilon^{A}_{2})} \, ,
$$
where  one recognizes the law for successive Poincar\'{e} transformations, $(\Lambda_{2},\epsilon^{A}_{2})$$(\Lambda_{1},\epsilon^{A}_{1})$ = $(\Lambda_{2}\Lambda_{1},\Lambda_{2}\epsilon^{A}_{1}+\epsilon^{A}_{2}).$

When just one transformation is applied to a plane wave, the phase factor at each event undergoes the same transformation. When there is a distinct transformation $(\Lambda,\epsilon^{A})$ at each event $A,$ a suitable generalization is to {\it{apply the distinct transformation at each event to the phase factor at that event.}} 

Therefore, the transformations $(\Lambda,\epsilon)$ are applied event-by-event to an initial plane wave $\exp{( i p_{0} \cdot y_{0})}.$ At event $A,$ this yields
  \begin{equation} \label{phase1}  e^{i p_{0} \cdot y^{A}_{0}} \rightarrow e^{i p \cdot (y^{A} +\epsilon^{A})} =  e^{i p \cdot y^{A}}e^{i p \cdot(- M + \Lambda^{-1} M \Lambda)\lambda y^{A}} e^{i p \cdot M_{0} \delta x}\, ,
\end{equation}
where the last expression follows from (\ref{epsilon1}), $p$ = $\Lambda p_{0},$ $y$ = $\Lambda y_{0},$ $M_{0}$ is the tensor in the initial reference frame and $M$ is in the transformed frame with $M_{0}$ = $\Lambda^{-1} M \Lambda $ and $M$ = $\Lambda M_{0} \Lambda^{-1}.$ 

Recall that $M$ is related to the curved spacetime metric $g_{\mu \nu},$ see (\ref{g3}). For a particle moving in weak gravitational fields, the gravitational field changes little over a region of space confining the particle for a short time. One can treat $M$ as constant over fairly large regions, large on the scale of the relevent portion of the quantum field. And on such a scale, the change of the momentum $p$ is often small due to a weak gravitational force. Thus we can treat both $M$ and $p$ as constants on a scale much larger than the scale of the quantum field. 

Then the coefficient of $y^{A}$ in the phase of the exponential in (\ref{phase1}) does not depend on the event $A.$ This means that a unique, common plane wave is formed by the process over the region where $M$ and $p$ can be considered constant. The momentum of the common plane wave is an `effective momentum' $\bar{p}$ given in
  $$  \bar{p} \cdot y = p \cdot y + p \cdot(- M + \Lambda^{-1} M \Lambda) \lambda y = p(1 - M \lambda + \Lambda^{-1} M \Lambda \lambda)\cdot y \, .
$$
The two momenta $\bar{p}$ and $p$ are equal when $\Lambda$ = 1 or $M$ = 1.

Keeping track of indices and displaying the flat spacetime metric $\eta$ yields an expression for the effective momentum. One has
  \begin{equation} \label{pbar}  \bar{p}^{\alpha} =  p^{\beta}\left[\delta^{\alpha}_{\beta} - \eta^{\alpha \rho}\eta_{\beta \tau}M^{\tau}_{\sigma}\lambda^{\sigma}_{\rho} + \eta^{\alpha \rho}\eta_{\beta \tau} (\Lambda^{-1} M \Lambda)^{\tau}_{\sigma} \lambda^{\sigma}_{\rho} \right] \, ,
\end{equation}
where summation over repeated indices is understood. Note that in general $\bar{p}$ is not the trajectory momentum $p$ = $\Lambda p_{0}$ that one would observe by measuring the time for a particle of mass $m$ to travel a known distance along the trajectory.  

Since the plane wave with the effective momentum $\bar{p}$ is the set of phase factors at each event in spacetime that describes the state of the particle, we conclude that {\it{for experiments that measure the momentum of cosmic rays by the transfer of momentum to other particles, it is the effective momentum $\bar{p}$  that is recorded.}}

By making suitable assumptions for $M,$ $\Lambda,$  and $\lambda$ we can apply the expression for the effective momentum (\ref{pbar}) to cosmic rays so that they have less effective energy traveling in interstellar space ($M$ = 1) than they have when detected in the gravitational potential of the Earth ($M \neq$ 1). 

We only consider protons as primaries. Consider a moving proton well-separated from other protons in some given reference frame. As in Sec. \ref{ClassLimit}, the relevant portion of the proton quantum field can be confined to a tube by combining particle states with momenta spread out over a suitable range to conform with the Heisenberg Uncertainty Principle.

Spacetime outside the tube is not be translated. We assume that the external spacetime remains in a given reference frame. We are most interested in ultrahigh energy cosmic rays, those with energies of $10^{18}$ eV or more, relative to the Earth and Sun. So we choose the given flat spacetime reference frame, the frame with events denoted $x$ above, to be a frame with the Earth and Sun moving with speeds negligible with respect to the speed of light. 

A suitably universal frame is provided by the distribution of Cosmic Microwave Background (CMB) radiation. The observed dipole anisotropy of the CMB implies the Solar System is moving at a speed of 370 km/s = 0.00123$c$ \cite{CMB1,CMB2} with respect to the CMB reference frame. We take the given frame to be the CMB reference frame.

Then the motion can be pictured as a succession of translations through displacements $\delta x_{i}$ applied to the region of spacetime inside the tube. Since the proton is confined to the tube, the resulting sequence of quantum fields inside the tube is just the same as if the translations were applied to the whole of spacetime. 

Since the motion of the proton is described as a sequence of translations with displacements in various spacetime directions, the accelerations of the proton are described by translations only, i.e. without any rotations or boosts. The parallel translation of the eigenmomentum allows and directs the proton's accelerations.  

Now there is a problem: By (\ref{pbar}), each effective momentum $\bar{p}_{i}$ depends on an as-yet-unspecified Lorentz transformation $\Lambda_{i}.$ Thus there is a sequence of some as-yet-unspecified initial reference frames that are translated to the given CMB frame with the as-yet unspecified transformations $\Lambda_{i}.$ 

We choose the $\Lambda_{i}$ so that the effective momenta $\bar{p}_{i}$ depend on the four-velocity of the proton in the CMB frame. In this way the effective energies that already depend on gravity can also depend on the proton's four-velocity. 

{\it{Assumption: The initial frame that determines each Lorentz transformation $\Lambda_{i}$ is the rest frame of the proton.}}

 Then each initial momentum ${p_{0}}_{i}$ is the momentum at rest, $k$ = $\{0,0,0,m\},$ and the Lorentz transformation $\Lambda_{i}$ is the transformation $L(p_{i}),$ Eq. (\ref{L}), taking $k$ to momentum of the proton's trajectory $p_{i}$ in the CMB reference frame. We have  
 \begin{equation} \label{Lp}\Lambda_{i}k = L(p_{i})k = p_{i} \, .
\end{equation}
From here on the sequence index is dropped, e.g. $\Lambda_{i} \rightarrow $ $\Lambda$ and $p_{i} \rightarrow$ $p,$ etc.

Having determined $\Lambda,$ we turn now to $M.$ For simplicity we assume a spherically symmetric, diagonal curved spacetime metric $g_{\mu \nu}.$ For trajectories in a weak gravitational field with the gravitational potential $\phi$, with both $\phi \leq$ 0 and $\mid \phi \mid <<$ 1, one has \cite{Adler}
\begin{equation} \label{gNEWT} g_{\mu \nu} = {\mathrm{diag}}(g_{xx},g_{xx},g_{xx},g_{tt}) = {\mathrm{diag}}(1 - 2 \phi,1 - 2 \phi,1 - 2 \phi, -1 - 2 \phi) \, .
\end{equation}
By (\ref{g3}), one choice for $M$ is diagonal,  
\begin{equation} \label{Mnewt} M^{\alpha}_{\mu} = {\mathrm{diag}}(M_{x},M_{x},M_{x},M_{t}) = {\mathrm{diag}}(1 -  \phi,1 -  \phi,1 -  \phi,1 -  \phi,1 +  \phi) \, ,
\end{equation}
where terms of second order in $\phi$ are dropped. 

 Referenced to a null potential at infinite distances from the sources, the weak field gravitational potential $\phi$ at an event $Q$ in space is the sum over sources, 
\begin{equation} \label{phi} \phi = - \sum_{s} \frac{ G m_{s}}{r_{s} c^{2}} \, ,
\end{equation}
where $G$ is the universal gravitational constant, $m_{s}$ is the mass of the source  $s,$ $r_{s}$ is the spatial distance from $s$ to the event $Q,$ and $c$ is the speed of light. 

Implicit in the application of Eqs. (\ref{Mnewt}) and (\ref{phi}) is the assumption that the potential of the galactic disk vanishes outside the immediate neighbohoods of stars. We make this assumption for simplicity and because it gives plausible results. 

Having determined $M$ and $\Lambda,$ it remains to consider $\lambda,$ which relates the reference frame for the trajectory and the frame for the effective particle states, $x$ = $\lambda y.$ If we assume the two frames are the same or opposite, i.e. $ \lambda$ = $\pm 1$ in (\ref{yTOx}) and (\ref{pbar}), then one finds that one can have the desired reduction in effective energy for the cosmic rays. But the effective three-momentum $\overrightarrow{\bar{p}},$ the spatial part of the four-momentum $\bar{p},$ would decrease with more energy, which is counter-intuitive. 

Heuristically, one can argue as follows. The lowest order term $M$ = 1 cancels out of (\ref{pbar}). The next order term has $-\phi$ for $M_{x}$ and $+\phi$ for $M_{t},$ see (\ref{Mnewt}), making this part of $M$ proportional to the matrix diag$\{-1,-1,-1,+1\}.$ It may be that the term $M \lambda$ is proportional to the identity (= diag$\{+1,+1,+1,+1\}$), since the first term in parentheses in (\ref{pbar}) is the identity. A spatial inversion or a time inversion for $\lambda$ combined with the part of $M$ linear in $\phi$ yields a matrix $M\lambda$ proportional to the identity, which can be expected to behave reasonably. One suspects that $\lambda$ could be an inversion. 
 
One finds that assuming $\lambda$ to be a time inversion produces a spatial momentum that increases when the energy increases, as one would expect. Thus, we assume that $\lambda$ {\it{is the time inversion}} 
\begin{equation} \label{lambda} \lambda^{\sigma}_{\rho} = {\mathrm{diag}}(+1,+1,+1,-1) \, .
\end{equation}
By the definition of $\lambda$ in (\ref{yTOx}), the reference frame for the particle states differs from the frame for the quantum field by a time inversion. This completes the collection of $M,$ $\Lambda,$ and $\lambda$ used to explain the cosmic ray spectrum.

By (\ref{pbar}), (\ref{Lp}), (\ref{Mnewt}), and (\ref{lambda}) we have the effective momentum 
\begin{equation} \label{pbar3} \bar{p}^{\mu} = p^{\mu} (1 - 4 \gamma^2 \phi)  \, ,
\end{equation}
where only the lowest order term in $\phi$ is kept,  $\bar{p}^{k}$ indicates the spatial part of the effective four-momentum, $k \in$ $\{1,2,3\},$ and $\bar{p}^{t}$ = $\bar{p}^{4}$ = $mc^2+\bar{E}$ is the effective total energy of the particle state, including the rest energy. Recall that $p$ (no bar) is the trajectory four-momentum observable by time-of-flight measurements.

By (\ref{pbar3}) the effective energy $\bar{p}^{t}$ is more than the energy of the trajectory since $\phi <$ 0. As argued previously, the measured cosmic ray energy is the quantity $\bar{E}$ = $\bar{p}^{t} - m c^{2}$ with the total energy $\bar{p}^{t}$ given in Eq. (\ref{pbar3}). One finds  
\begin{equation} \label{energy}
\bar{E} = E(1 - 4  \gamma^{2} \phi) \, .
\end{equation}
For the accuracy needed here, the thickness of the atmosphere can be neglected.  By (\ref{phi}), we find the gravitational potential $\phi$ to be $-1.06 \times 10^{-8}$ at the Earth's surface due to the Sun and the Earth.

Experiments that measure the energy of cosmic rays have succeeded in pushing the observed spectrum to `ultra-high' energies $E > 10^{18}$ eV. Some of the data \cite{data1} - \cite{data10} is collected in the spectrum labeled `Earth' in Fig. 2. Not all available data is included because the discussion here involves the coarse properties of the spectrum upon which all data agree. 

The expected energy spectrum of the protons in interstellar space (I.S.) can be obtained from the energies of cosmic rays observed by Earth-based experiments. One applies the chain rule to obtain the predicted scaled flux in interstellar space using (\ref{energy}),
\begin{equation} \label{chain} \biggl(\frac{dN}{dlnE}\biggr)_{I.S.} =  \biggl(\frac{dN}{dln\bar{E}}\biggr)_{Earth} \, \frac{dln\bar{E}}{dlnE} \, ,
\end{equation}
where $(dN/dln\bar{E})_{Earth}$ is the scaled flux observed by Earth-bound experiments, $\bar{E}$ = $\bar{p}^{t} - m c^{2}$ and ${E}$ = ${p}^{t} - m c^{2}.$
The resulting energy spectrum is labelled `Interstellar Space' in Fig. 2. 

With (\ref{energy}) and (\ref{chain}) one can derive the energy spectrum in interstellar space from the observed Earth-based spectrum. Now reverse the process and predict the cosmic ray spectrum at the surface of the Sun where $\phi$ = $-2.12 \times 10^{-6},$ with $\phi$ = 0 in interstellar space as before. The spectrum is labelled `Sun' in Fig. 2. 

One sees from Fig. 2 that the three spectra overlap up to about $10^{13}$ eV, where the spectra diverge and at the highest energies the spectra differ by many orders of magnitude. The predicted interstellar spectrum may be compared with the astrophysics of accelerating and transporting the protons. Such a comparison lies beyond the scope of this paper. It may be possible to detect cosmic rays striking the Sun. If so, then the ultrahigh cosmic rays striking the Sun should have energies more than a hundred times larger than the same cosmic rays incident on the Earth.  

Confirmation of the predicted cosmic ray spectra in interstellar space and at the Sun's surface would provide support for the assumptions and explanation presented in this section. 

There is some experimental evidence that does not favor the explanation proposed here. If the trajectory energies of cosmic rays protons are as low as expected here, only up to about $10^{15}$ eV, then galactic magnetic fields should randomize their trajectories in interstellar space. Conversely, the Auger collaboration finds a nonisotropic distribution of the incident direction of the highest energy cosmic rays.\cite{Auger} Another consequence of low energy ($10^{15}$~eV) trajectories would be the lack of a GZK cutoff, since the cutoff takes effect at $10^{19}$~eV. However such a cutoff is reported by the HiRes experiment.\cite{HiRes} These experiments are ongoing and difficult and the reports are still controversial. A definitive test of the explanation proposed here is expected when the Large Hadron Collider begins exploring the energy range where the spectra in Fig. 2 diverge.


\begin{figure}[ht] \label{f1} 
\centering
\vspace{0cm}
\hspace{0in}\includegraphics[0,0][360,360]{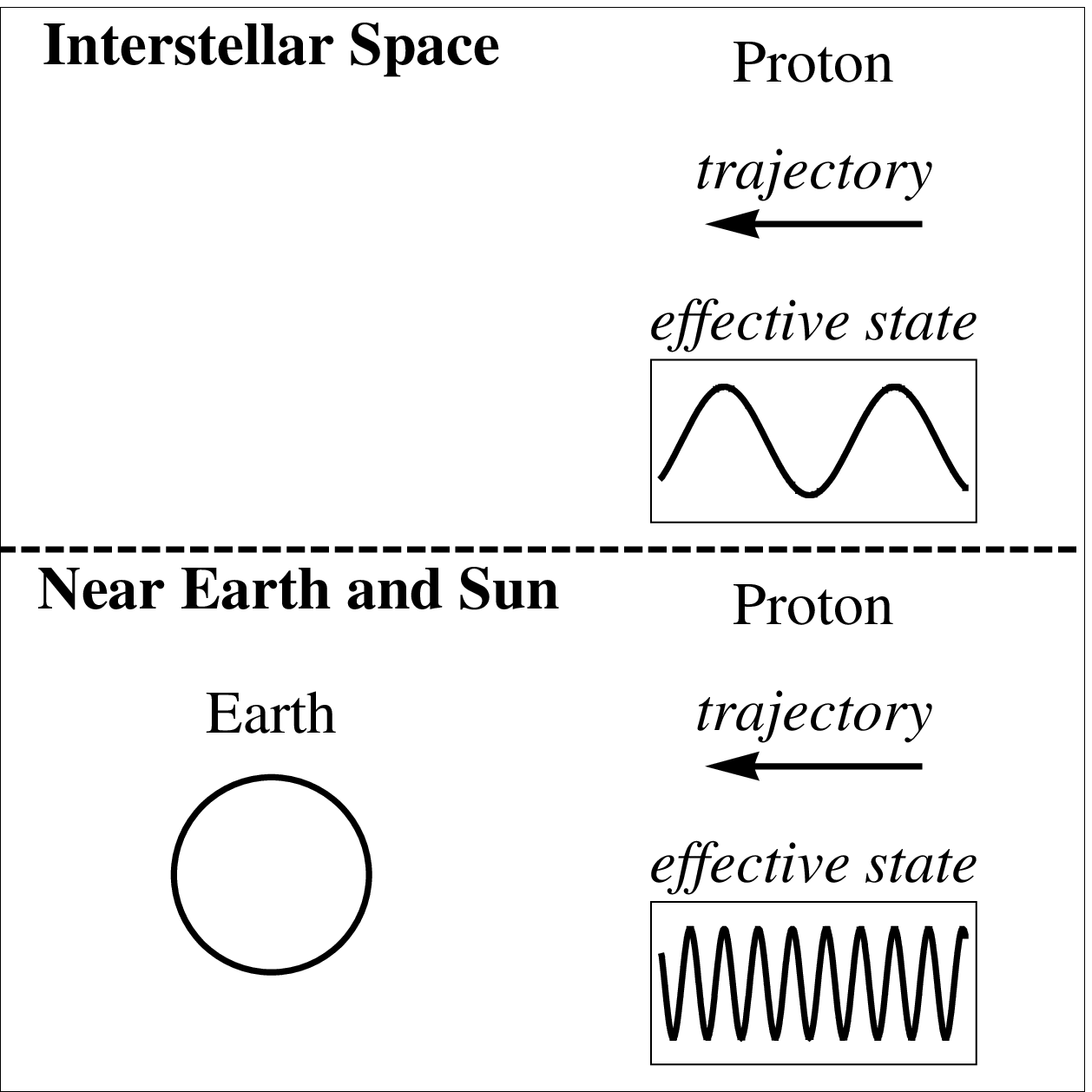}
\caption{{\it{An Explanation for Ultrahigh Cosmic Ray Energies.}} In this explanation of cosmic ray energies, the constraints on the quantum field allows a proton or other massive particle to have two energies. One is the trajectory energy measured by mass and time-of-flight. The other is the effective particle state energy evident in state-changing collisions. Top: In interstellar space, the trajectory energy and the effective state energy are equal. Bottom: In the gravitational field of the Earth and the Sun, conventional gravity has increased the trajectory energy, but only negligibly. But the effective state energy is some $10^5$ times higher than the trajectory energy. It is the ultrahigh effective  state energy that is deposited into the Earth's atmosphere when the effective particle state changes upon interaction with the atmosphere. }
\end{figure}

\begin{figure}[ht] \label{f2}	
\centering
\vspace{0cm}
\hspace{0in}\includegraphics[0,0][360,360]{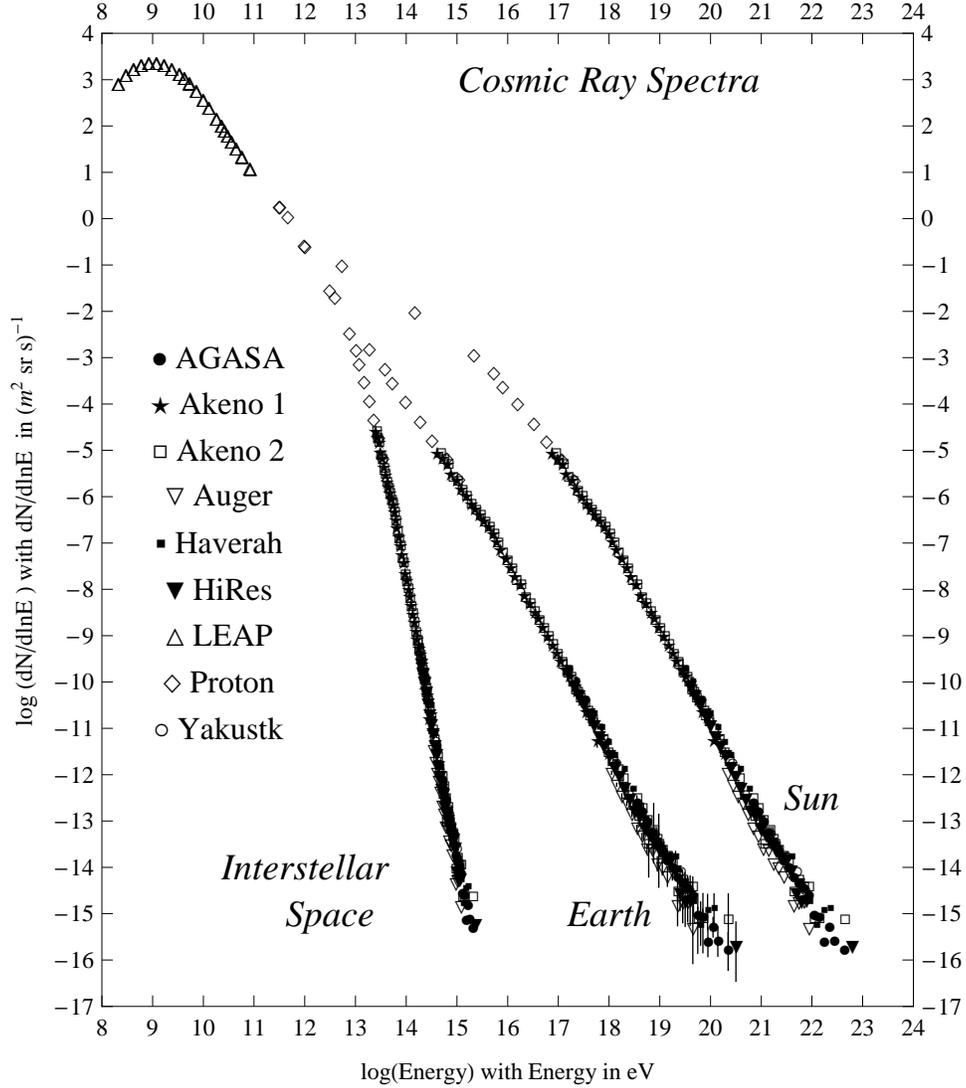}
\caption{{\it{Cosmic Ray Spectra, Dependence on Local Gravity.}} The spectrum labeled `Earth' plots the experimentally determined scaled flux incident on Earth's atmosphere.\cite{data1}-\cite{data10} The flux is scaled by the trajectory kinetic energy $E,$ i.e. $dN/dlnE$ = $E dN/dE.$ By Eq.~(\ref{energy}), the spectrum `Interstellar Space' is the expected spectrum where the gravitational potential vanishes. For the most energetic cosmic rays, the predicted Interstellar Space spectrum is remarkably less energetic than the spectrum observed on Earth. At the Sun the magnitude of the gravitational potential is greater than at the Earth and the predicted cosmic ray energies at the Sun are higher than those detected at the Earth.} 
\end{figure}

\begin{figure}[tbp] 
\centering
\vspace{0cm}
\hspace{0in}\includegraphics[0,0][360,253]{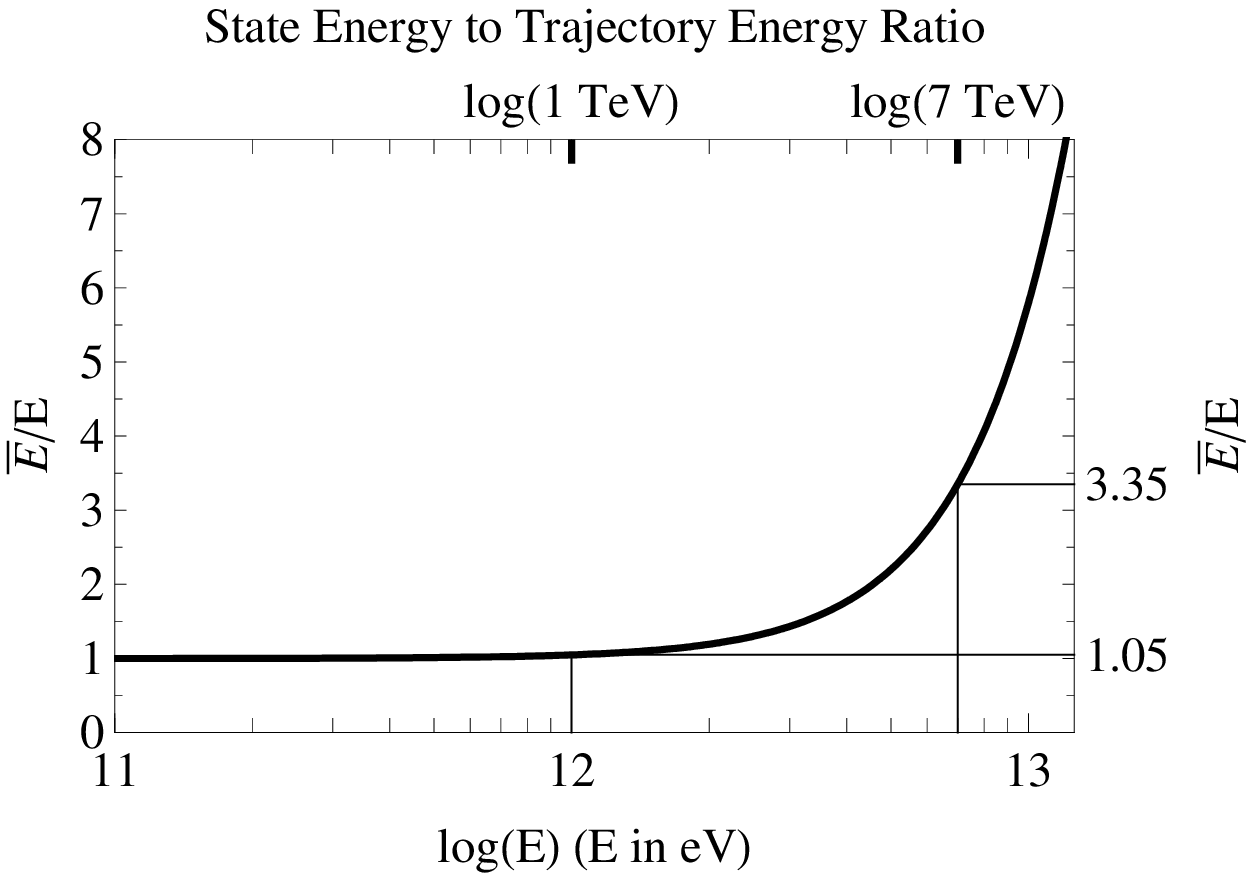}
\caption{{\it{Testing the Explanation at the LHC.}} The explanation of cosmic rays gives a proton or other massive particle two energies whose ratio is $\bar{E}/E$ =  $1-4 \phi \gamma^{2},$ where $\phi$ is the gravitational potential on Earth divided by the square of the speed of light and where $\gamma$ is the relativistic gamma of the proton's trajectory. See Eq.~(\ref{energy}). Given 7 TeV as a proton trajectory energy $E,$ collisions should deliver a particle state energy $\bar{E}$ per proton of 23.5 TeV (= $3.35 \times 7$ ). Observation at the LHC of a $2 \times 23.5$ = 47 TeV collision from an antiproton and proton each with 7 TeV of trajectory energy would confirm the cosmic ray explanation proposed here. }
  \label{fig:fig1b}
\end{figure}

\vspace{1.0cm}

\section{Accelerator Experiments} \label{LHC}

A strong test of the proposed effects is expected when the Large Hadron Collider (LHC) \cite{LHC} becomes operational. The LHC is designed to operate with protons of up to 7 TeV, just where the proposed effect begins to show up in the cosmic ray spectra of Fig. 2. 

The LHC is designed to collide proton and antiproton beams. The LAB frame is the same as the Center-of-Mass (CM) frame since in the LAB frame the protons and antiprotons approach each other with the same energy. And, for the accuracy needed here, we can ignore the relative velocity of the LAB frame with the cosmic microwave background CMB frame. Thus, for the purposes here, the LAB and CM frames at the LHC are the same as the CMB frame chosen as the given frame in Sec. \ref{s3} for cosmic rays.

By the explanation of ultrahigh energy cosmic rays presented in Sec. \ref{s3}, we expect that the trajectory momentum $p$ and particle state momentum $\bar{p}$ to be related as in Eq. (\ref{pbar3}). The trajectory momentum $p$ is the one that determines deflections in the electromagnetic fields of the LHC, as is expected by Eq. (\ref{D2x3}) and Eq. (\ref{D2x4}).  The particle state momentum $\bar{p}$ is the momentum measured in collisions. 

Thus, the trajectory energy $E$ should be less than the particle state energy $\bar{E}$ as given by the formula (\ref{energy}). See Fig. 3 for a graph of the ratio of these two energies in the energy region of the proton and antiproton beams at the LHC. 

If the suppositions in this article are correct, then the proton bunches move through the various electromagnetic fields with up to 7 TeV trajectories, in keeping with the design. Since an $E$ = 7 TeV proton trajectory energy has a gamma of $\gamma \approx$ 7500, we find by Eq.~(\ref{energy}) that the proton particle state energy should be 23.5 TeV.  Thus when a proton collides with an antiproton the collision energy is not twice the trajectory energy, i.e. $2 \times 7$ = 14~TeV, but twice the particle state energy, i.e. $2 \times 23.5$ = 47~TeV. Thus the difference between the trajectory and effective energies predicted in this article should be grossly evident to the LHC experiments' calorimeters. 

By comparison, an $E$ = 1 TeV trajectory proton at the Tevatron \cite{Tevatron} would have a particle state energy only 5\% more than the trajectory energy and deliver 1.05 TeV per proton, not much more than expected. It is a coincidence of the value of the Earth's gravitational potential and the new range of the proton's energy from 1 TeV to 7 TeV expected at the LHC that the relevent term in Eq. (\ref{energy}), the term $4 \gamma^{2} \phi,$ becomes significant.  The first high energy runs at the LHC should make it obvious whether or not the new effect proposed in Sec. \ref{s3} could be realized in nature.


\end{document}